\documentclass[%
 preprint,
 amsmath,amssymb,
 aps, physrev,
]{revtex4-2}

\usepackage{graphicx}
\usepackage{dcolumn}
\usepackage{bm}

\begin{document}

\preprint{APS/123-QED}

\title{\textbf{Direct observation of reversible ionic polarization at a buried solid-state battery interface.} 
}%

\author{Giovanni Ceccio}
\email{ceccio@ujf.cas.cz}
\author{Romana Mikšová}%
\author{Jiri Vacik}%
\affiliation{
 Nuclear Physics Institute of Czech Academy of Sciences, Department of Neutron and Ion Methods, Husinec-Řež, Czech Republic
}%

\author{Zoltan Szaraz}
\email{zoltan.szaraz@stuba.sk}
\author{Josef Dobrovodský}%
\author{Pavlina Zavadilová}
\affiliation{
 Advanced Technologies Research Institute, Faculty of Materials Science and Technology in Trnava, Slovak University of Technology in Bratislava, Trnava, Slovakia
}

\author{Pavol Noga}
\affiliation{%
  Advanced Technologies Research Institute, Faculty of Materials Science and Technology in Trnava, Slovak University of Technology in Bratislava, Trnava, Slovakia
}%
\affiliation{
Centre for Interdisciplinary Research (CIDR), SRM University-AP, Amaravati, Andhra Pradesh, India[
}%
\author{Ivan Mastronardo}
\affiliation{%
  CNR-ITAE, salita Santa Lucia Sopra Contesse, 5, Messina (ME) 98126, Italy
}%
\affiliation{
Department of Engineering, University of Messina, Contrada di Dio, Messina 98158, Italy
}%

\date{\today}

\begin{abstract}
Electric fields control ionic distributions at solid-solid interfaces. However, whether buried interfaces possess an intrinsic ionic reversibility has remained experimentally unresolved. Here, we directly observe reversible ionic polarization at the buried amorphous LiCoO$_2$/Li$_1.5$Al$_0.5$Ge$_1.5$P$_3$O$_{12}$ electrolyte interface. By combining the Elastic Recoil Detection Analysis (ERDA) method, utilizing time-of-flight (ToF) measurements and nanometer-scale resolution, with the Neutron Depth Profiling (NDP) method for the absolute quantification of lithium, we show that an applied bias of ±1 V results in a reversible redistribution of approximately 9.3$\%$ of lithium in the interface region, while conserving its total content. Upon polarity reversal, the lithium redistribution is completely reversed, indicating reversible interfacial ionic polarization rather than irreversible lithium loss or permanent interphase formation. 
These results demonstrate that buried interfaces in solid-electrolyte systems can behave as electrically reconfigurable ionic regions with an intrinsic reversible freedom for ion motion, thereby providing direct microscopic insight into the physics of space-charge formation and low-voltage interfacial response in solid-state electrochemical systems.

\end{abstract}

\maketitle


Electric fields redistribute mobile ions at solid-solid interfaces, thereby modifying local electrochemical potentials, charge distributions, and ionic transport. While these processes are well understood in liquid electrolytes, their microscopic manifestation at buried solid-state interfaces remains largely elusive, as the associated lithium distributions cannot be directly observed. Consequently, the question regarding the   physical nature of a buried solid-solid interface under an applied electric field remains unresolved. Do these interfaces act merely as passive geometrical boundaries, undergo irreversible changes due to the interphase formation, or are they capable of exhibiting an intrinsic, reversible ionic response? Thin-film all-solid-state batteries serve an ideal model system for investigating this phenomenon issue because their active layers are only tens to hundreds of nanometers thick; this means that even small displacements of lithium represent a significant fraction of the total ionic content \cite{son2025overcoming,zhang2021advanced, son2025five,amiki2013electrochemical,sun2017recent}. 
Theory predicts the formation of electrostatic potentials and space-charge layer at interfaces in solid-state battery interfaces \cite{swift2019first}. Earlier studies suggested that nanoscale interfaces can exhibit properties representing a transition between capacitive and battery-like storage modes, where space charge, electrochemical potential, and interfacial stability are key physical quantities \cite{zhukovskii2006evidence}. However, no direct experimental evidence has yet been found showing that buried solid-state interfaces can reversibly redistribute lithium under the influence of an electric field. The main challenge lies in the fact that lithium is both light and located buried within the material's structure. While methods utilizing electrons or X-rays provide valuable structural or chemical information, they do not allow for quantitative determination of lithium and its distribution, particularly in multilayer thin film structures. Thanks to the $^6$Li(n,$\alpha$)$^3$H nuclear reaction, Neutron Depth Profiling (NDP) is a method exceptionally sensitive to lithium and its results are independent of the chemical state-or phase, making it a powerful tool for the absolute quantification of lithium \cite{liu2014situ,fuller2021spatially,nagpure2011neutron,tomandl2020analysis,pivarnikova2024observation}. 
However, for ultrathin multilayers, its depth resolution (few tens of nanometers) is insufficient on the nanometer scale within the buried interface region between the cathode and the electrolyte. 
In contrast, the ToF-ERDA technique (Time-of-Flight Elastic Recoil Detection Analysis) enables the acquisition of quantitative multi-element depth profiles with high sensitivity to lithium and nanometer-scale resolution \cite{mathayan2021assessing,hong1997development, dobrovodsky2026high, noga2017new}. In this work, we combine both methods: NDP to determine the total lithium content, and high-resolution ToF-ERDA, which allows for the quantitative profiling of light elements such as Li, and the determination of the lithium's position within the thin multilayer system. 
The investigated multilayers were deposited on polished Al$_2$O$_3$ polycrystalline ceramic substrates using ion-beam sputtering at the LEIF (Low Energy Ion Facility) at Nuclear Physics Institute, Řež \cite{ceccio2025study, ceccio2026hybrid}. 
The multilayer architecture consisted of an Al$_2$O$_3$/Cu/Fe$_2$O$_3$/LAGP/LCO/Cu arrangement, with Fe$_2$O$_3$ serving as the anode layer, Li$_1.5$Al$_0.5$Ge$_1.5$P$_3$O$_{12}$ (LAGP) as the solid electrolyte, LiCoO$_2$ (LCO) as the cathode, and Cu layers as current collectors, Fig. 1. LAGP was selected as the solid electrolyte due to its good stability in air and wide electrochemical stability window, which enables reliable processing, particularly in combination with LCO \cite{cretu2023impact, jung2025calcination, park2023co, li2024stabilizing} . 

\begin{figure}[h]
\includegraphics[width=0.8\textwidth]{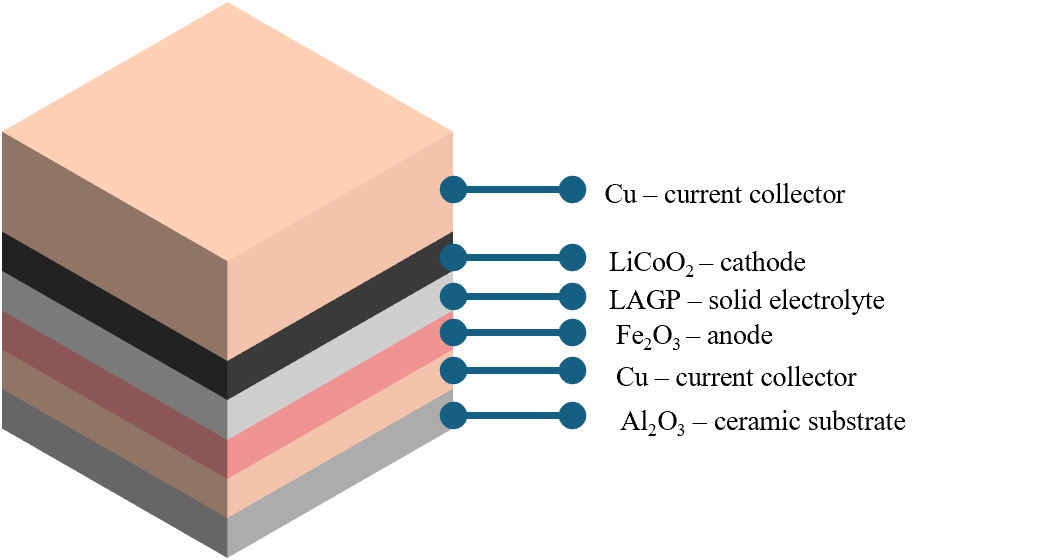}
\caption{\label{fig1}Schematic of the Al$_2$O$_3$/Cu/Fe$_2$O$_3$/LAGP/LCO/Cu thin-film battery.}
\end{figure}

The layers were deposited at room temperature to minimize thermally activated lithium diffusion during growth, ensuring the formation of an amorphous thin multilayer. The total thickness of this stack structure was approximately 450 nm. 
This amorphous architecture provides  a simplified model system for studying the response of lithium to an applied electric field at low voltage across the buried oxide cathode / electrolyte interface, minimizing with the influence of microstructural factors such as  grain- boundaries, crystallographic texture, and thermally induced crystallization.
Prior to the application of an electric voltage, the total lithium content was measured by NDP facility at the Nuclear Physics Institute (NPI). The NDP spectrometer is installed at the HK3 horizontal channel of the LVR-15 research reactor and utilizes a thermal neutron flux of 6 $\times$ 10$^7$ $n$ $cm^{-2}s^{-1}$. The NDP spectrum, calibrated using a certified boron reference standard (NIST Standard Reference Material 2137), yielded a total lithium areal density of $2.21$ $\pm$ 0.08 $\times$ 10$^{16}$ atoms cm$^{-2}$. Owing to its non-destructive nature, NDP provides an  absolute reference for the total lithium inventory prior to subsequent ion-beam depth profiling. The same multilayer structure was subsequently characterized in its pristine state  using the ToF-ERDA setup coupled to the 6 MV tandem accelerator of the Advanced Technology Research Institute MTF STU in Trnava \cite{noga2017new}. A 45 MeV iodine beam was used to determine the initial depth distribution of lithium within the LCO/LAGP layer system. The scattering angle between the incident beam and the ToF-ERDA telescope axis is 40.6$^{\circ}$, corresponding to standard incident and exit  angles of 20.3$^{\circ}$ relative to the sample surface.  The resulting elemental profiles shown in Fig. 2 clearly resolve the multilayer architecture: Cu identifies the current collectors, Co the LCO cathode layer, P, Ge, and Al the LAGP electrolyte. Lithium is concentrated primarily at the cathode/electrolyte interface; in the as-deposited state, a measurable signal extends into the adjacent Fe$_2$O$_3$ layer.  The corresponding lithium distributions are shown in the ToF-ERDA profiles in Fig. 3. 
\begin{figure}[h]
\includegraphics[width=0.8\textwidth]{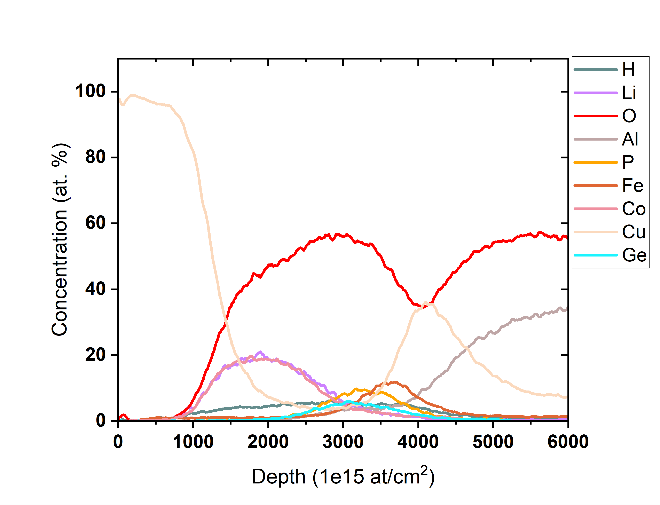}
\caption{\label{fig2}Elemental profile from the ToF-ERDA analysis, the two Cu peaks delimit the sample multilayer structure.}
\end{figure}

\begin{figure}[h]
\includegraphics[width=0.8\textwidth]{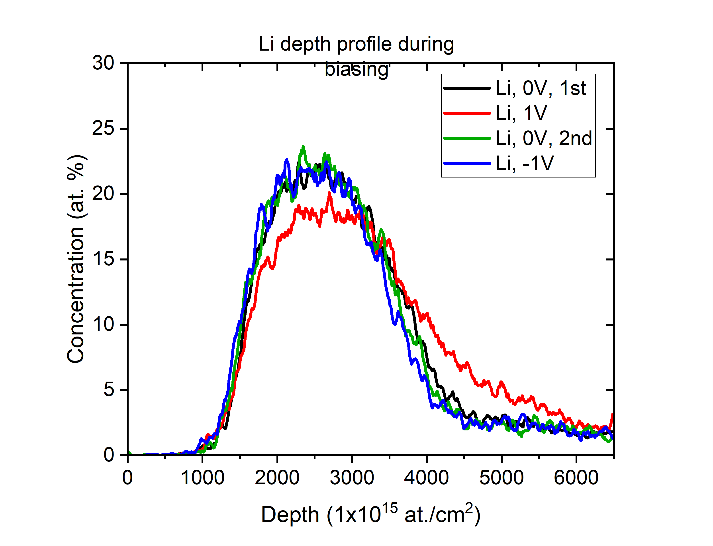}
\caption{\label{fig3}Lithium depth profile for several biasing conditions, showing the reversible migration of Li ions between the cathode and the electrolyte.}
\end{figure}

The application of a potential of +1 V to the anode results in a decrease in lithium concentration within the cathode, accompanied by lithium accumulation at the adjacent LCO/LAGP interface on the electrolyte side. This behavior does not correspond to a uniform shift of the entire lithium profile; rather, it indicates a localized redistribution of a limited amount of lithium from the cathode toward the LCO/LAGP interface.
Integration of the lithium signal across the LCO and LAGP layers shows that the total lithium content remains constant within experimental uncertainty. The applied electric field therefore does not extract lithium from the analyzed structure or induce measurable lithium losses into the current collector or substrate. Instead, it drives only an internal redistribution of lithium. Quantitatively, the redistributed lithium corresponds to approximately 9.3\% of the total lithium content in the system. Based on the lithium areal density calibrated by the NDP method, this fraction represents 2.06 $\times$ 10$^{15}$ Li atoms per cm$^{-2}$. For the 5 mm$^2$ area analyzed by ToF-ERDA, this corresponds to approximately $1 \times 10^{14}$ lithium atoms, equivalent to a reversible ionic charge of 1.7 $\times 10^{-5}$ C. This value is physically significant because it demonstrates that the response to the electric field does not involve the migration of the entire lithium inventory within the layer, but rather a localized redistribution of a small fraction of lithium. 
Instead, the interface selects only a limited population of mobile lithium ions, while the majority of lithium remains immobile on the timescale of the experiment and under the applied low-voltage conditions. The dependence on polarity demonstrates the reversibility of this response. Upon removal of the positive bias, the lithium distribution returns close to its original state. Application of $-$1 V restores the lithium concentration within the cathode. The recovery of the original lithium distribution excludes irreversible lithium trapping at the interface and suggests that the observed change results from reversible internal lithium redistribution rather than decomposition, permanent cathode delithiation, or beam-induced migration or loss. This behavior is consistent with reversible ionic polarization, in which the applied electric field modifies the lithium electrochemical potential at the buried LCO/LAGP interface and drives the redistribution of a limited population of mobile lithium ions over a nanometer-scale region, while the overall lithium content remains unchanged.ToF-ERDA measurements show that this reversible redistribution corresponds to approximately 9.3$\%$ of the total lithium content at an applied voltage of $\pm$1 V. This fraction represents the experimentally observed amplitude of the reversible ionic polarization. The microscopic origin of this finite response remains unresolved and may reflect the combined effects of interfacial electrochemical equilibrium, structural disorder, heterogeneous migration landscapes, and space-charge phenomena. Nevertheless, complete reversibility indicates that the buried LCO/LAGP interface functions as a dynamically responsive ionic system rather than undergoing irreversible lithium consumption or permanent interphase growth.
The observation is distinct from the conventional descriptions of cathode / electrolyte interphase formation. An interphase implies the emergence of a chemically or structurally distinct layer at the interface. Our measurements do not require such an interpretation. Instead, they show that the existing LCO/LAGP interface supports a reversible lithium polarization profile under low bias conditions. Complementary operando structural and spectroscopic measurements will be required to determine whether prolonged cycling or operation at higher voltages eventually leads to the formation of a chemically distinct interphase. The amorphous nature of the sputtered layers also may play an important role. Unlike crystalline electrodes and solid electrolytes, where lithium transport is governed by well-defined crystallographic pathways, amorphous LCO/LAGP lacks long-range structural periodicity and instead presents a heterogeneous lithium energy landscape, which is characterized by a broad distribution of non-equivalent lithium sites with varying site energies and migration barriers \cite{schafer2018site}.
Under low applied voltage, the reversible response is expected to involve only lithium ions occupying energetically accessible sites within the heterogeneous interfacial energy landscape. Lithium ions associated with shallow potential wells or well-connected migration pathways can respond to the applied field, whereas those trapped in deeper energy minima remain immobile on the experimental timescale. This interpretation is consistent with the observation that only a limited fraction  of the lithium undergoes redistribution. 
At low bias, only the lithium located in sites with sufficiently weak bonding or sites associated with the phase interface reacts reversibly. This interpretation connects the experiment to a broader question in nonequilibrium ionics: how disorder,  interfacial electrochemical potential, and space-charge formation determine which ions become mobile under weak driving fields.

The complementary NDP/ToF-ERDA approach is critical for establishing the origin of the observed lithium redistribution. ToF-ERDA provides quantitative, depth-resolved lithium concentrations across the multilayer stack, but the approximately 5 mm² ion beam probes only a limited area of the battery surface. In contrast, NDP provides an integral measure of the lithium inventory across the entire battery area, but lacks the depth resolution required to resolve nanometer-scale redistribution within the approximately 450 nm-thick multilayer stack. The combination of these complementary measurements therefore links the locally resolved changes in the lithium depth profile to the independently determined total lithium inventory. The conservation of the total lithium content, together with the reversible changes observed in the depth profiles, demonstrates that the bias-induced response originates from internal lithium redistribution. This provides direct evidence for an interfacial ionic polarization process.

The result suggests a revised understanding of the interface between the cathode-solid electrolyte interface in thin-film batteries. This interface is not merely a structural junction or a chemically evolving degradation zone, it can actually function as a layer enabling reversible ionic polarization.
Under low-voltage conditions, this interfacial region accommodates a small but measurable amount of ionic charge, which is released upon reversal of the applied polarity. Such behavior may contribute to interfacial impedance, apparent capacitance, and the voltage response during the initial phase of operation, prior to the onset of irreversible degradation processes. Because of the limited amount of lithium involved in this process, this reversible ionic response may represent an important transition between equilibrium polarization and irreversible electrochemical transformation.
In conclusion, we observe a reversible electric-field-driven lithium polarization at the buried amorphous LCO/LAGP interface of a solid-state battery. Under an applied voltage of $\pm$1 V, approximately 9.3\% of the total lithium content was reversibly redistributed between the cathode region and the adjacent electrolyte interface, without any measurable change in the overall lithium content. The interface, therefore, behaves more like an electrically reconfigurable ionic region than merely a passive material interface or an irreversibly growing interphase. This finding provides direct microscopic evidence for reversible interfacial ionic polarization at the interface of a thin-layer all-solid-state battery and establishes a way for connecting nanoscale lithium redistribution with the space-charge effects, interfacial resistance, and low-voltage electrochemical response in solid-state ionic devices.

\begin{acknowledgments}
 The authors acknowledge the support provided by the Ferroic Multifunctionalities project, supported by the Ministry of Education, Youth, and Sports of the Czech Republic; Project No. CZ.02.01.01/00/22\_008/0004591, co-funded by the European Union, by the EU NextGenerationEU through the Recovery and Resilience Plan for Slovakia under project No. 09I04-03-V02-00046, and the Grant agency VEGA, project No. 1/0558/24.
\end{acknowledgments}

\bibliography{apssamp}

\end{document}